\documentstyle[12pt]{article}
\begin{document}
\setlength{\oddsidemargin}{-0.1cm}
\setlength{\topmargin}{-1.cm}

\title{The solution of multi-scale partial differential equations using wavelets}
\author{
Stefan Goedecker,
\\Max-Planck Institute for Solid State Research, 
\\Stuttgart, Germany
\\goedeck@prr.mpi-stuttgart.mpg.de
\\ Oleg Ivanov
\\P.N. Lebedev Physical Institute, Moscow, Russia
\\ivanov@td.lpi.ac.ru}
\maketitle
\section{Introduction}
Wavelets are a powerful new mathematical tool which offers the possibility 
to treat in a natural way quantities characterized by several length scales. 
In this article we will show how wavelets can be used to solve partial differential 
equations which are characterized by widely varying length scales and which are 
therefore hardly accessible by other numerical methods.
The standard way to solve partial differential equations is to express the solution 
as a linear combination of so-called basis functions. These basis functions can for instance be 
plane waves, Gaussians or finite elements. Having discretized the differential equation 
in this way makes it amenable to a numerical solution. 
Wavelets are just another basis set which however offers considerable advantages 
over alternative basis sets. Its main advantages are:
\begin{enumerate}

\item The basis set can be improved in a systematic way: \newline
 If one wants the solution of the differential equation with higher accuracy one 
 can just add more wavelets in the expansion of the solution. This will not 
 lead to any numerical instabilities.

\item Different resolutions can be used in different regions of space: \newline
 If the solution of the differential equation is varying particularly rapidly 
 in a particular region of space one can increase the resolution in this region by 
 adding more high resolution wavelets centered around this region. 

\item There are few topological constraints for increased resolution regions: \newline
 The regions of increased resolution can be chosen in arbitrarily, the only requirement being 
 that a region of higher resolution be contained in a region of the next lower resolution.
 
\item The matrix elements of the differential operators are very easy to calculate

\item The numerical effort scales linearly with respect to system size: \newline
 Three-dimensional problems of realistic size require usually a very large number 
 of basis functions. It is therefore of utmost importance, that the numerical 
 effort scales only linearly (and not quadratically or cubically) with respect 
 to the number of basis functions. If one uses iterative matrix techniques, this 
 requirement is equivalent to the two requirements, namely that the matrix vector 
 multiplications which are necessary for all iterative methods can be 
 done with linear scaling and that the number of matrix vector multiplications 
 is independent of the problem size. 
 The first requirement is fulfilled since the matrix 
 representing the differential operator is sparse. 
 The second requirement is related to the 
 availability of a good preconditioning scheme which can be easily found 
 by analyzing the Fourier properties of wavelets. 

\end{enumerate}

\section{A first tour of some wavelet families}
\begin{figure}[ht]    
     \begin{center}
      \setlength{\unitlength}{1cm}
       \begin{picture}( 8.,3.5)           
        \put(-1.,-1.5){\includegraphics{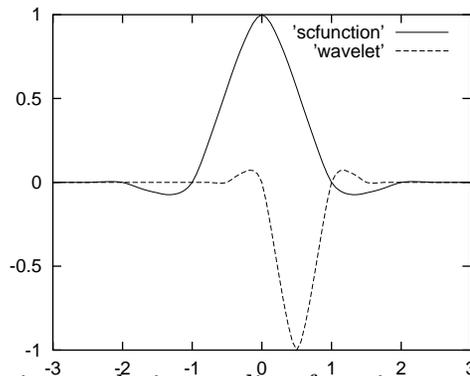}}   
       \end{picture}
       \caption{ \label{lazy4} The interpolating scaling function and wavelet of degree 4}
      \end{center}
     \end{figure}

Many families of wavelets have been proposed in the mathematical literature.
If one wants to use wavelets for the solution of differential equations, one 
therefore has to choose one specific family which is most advantageous for the 
intended application. Within one family there are also 
members of different degree. We believe that the so-called bi-orthogonal 
interpolating wavelets~\cite{lazy} are the most useful ones in the context of 
differential equations and we will therefore mainly concentrate on this class.
Each wavelet family 
is characterized by two functions, the mother scaling function $\phi$ and the 
mother wavelet $\psi$. For the case of a fourth order interpolating wavelet they are 
shown in Figure~\ref{lazy4}.

\begin{figure}[ht]       
     \begin{center}
      \setlength{\unitlength}{1cm}
       \begin{picture}( 8.,3.)           
        \put(-2.,-.5){\includegraphics{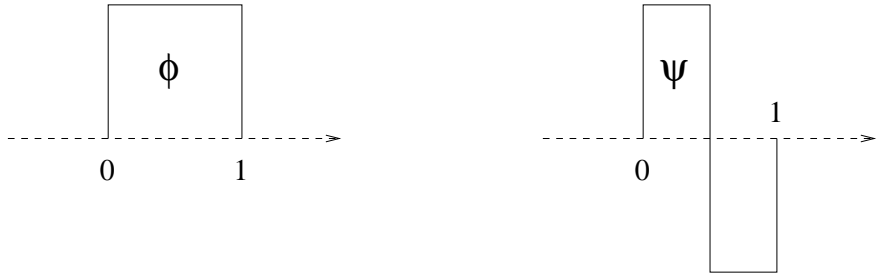}}   
       \end{picture}
       \caption{\label{haarfam} The Haar scaling function and wavelet.}
      \end{center}
     \end{figure}

Another family which will be introduced is the 
Haar wavelet family shown in Figure~\ref{haarfam}. 
It is too crude to be useful for any 
numerical work, but its simplicity will help us to illustrate some basic 
wavelet concepts.

To obtain a basis set at a certain resolution level $k$ one can use all the 
integer translations of the mother scaling function of some wavelet family.

\begin{equation} \label{sf2ind}
\phi_{i}^{k}(x)  = \phi( 2^k x-i)
\end{equation}

Note that with this convention higher 
resolution corresponds to larger values of $k$. 
Exactly the same scaling and shifting operations can of course also be applied to the 
wavelets.
\begin{equation} \label{wv2ind} 
\psi_{i}^{k}(x)  = \psi( 2^k x-i)
\end{equation}
This set of wavelet  basis functions can be added as a basis to the scaling functions 
as will be explained in the following for the case of the Haar wavelet family.

\section{The Haar wavelet}
\begin{figure}[th]          
     \begin{center}
      \setlength{\unitlength}{1cm}
       \begin{picture}( 8.,2.5)           
        \put(-1.,-.5){\includegraphics{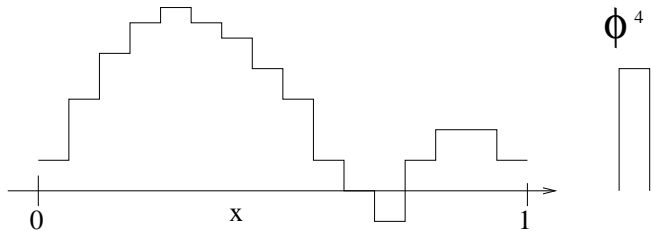}}   
       \end{picture}
       \caption{\label{fhaarm4} function at resolution level 4.}
      \end{center}
     \end{figure}

In the case of the Haar family, any function which can exactly be represented at 
any level of resolution is necessarily piecewise constant. One such function 
is shown in Figure~\ref{fhaarm4}.

 Evidently this 
function can be written as a linear combination of the scaling functions 
$\phi_{i}^{4}(x)$

\begin{equation} \label{hsfcrep}
f = \sum_{i=0}^{16} s_{i}^{4} \: \phi_{i}^{4}(x) 
\end{equation} 

where $s_{i}^{4} = f(i/16) $.

\begin{figure}[ht]            
     \begin{center}
      \setlength{\unitlength}{1cm}
       \begin{picture}( 8.,7.5)           
        \put(0.,-.8){\includegraphics{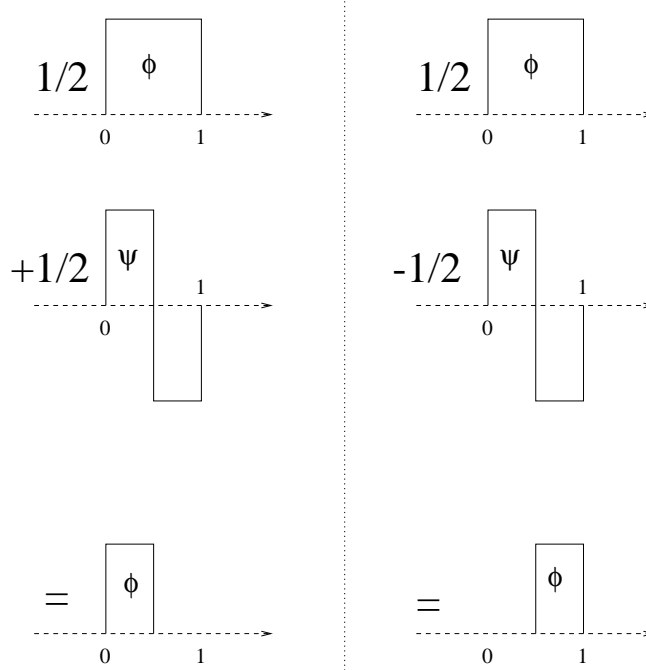}}   
       \end{picture}
       \caption{\label{haarcombo} A skinny scaling function is a linear combination of 
       a fat scaling function and a wavelet.}
      \end{center}
     \end{figure}

Another, more interesting, possibility consists of expanding a function with respect 
to wavelets of different resolution. This is possible because a scaling 
function (and wavelet)  at resolution 
level $k$ is always a linear combination of a scaling function and a wavelet at the 
next coarser level $k-1$ as shown in Figure~\ref{haarcombo}

Using this relation, we can write any linear combination of the two 
scaling functions $\phi_{2 i}^{k}(x)$ and $\phi_{2 i+1}^{k}(x)$ as a linear combination of 
$\phi_{i}^{k-1}(x)$ and $\psi_{i}^{k-1}(x)$.

Denoting the expansion coefficients with respect 
to $\psi_{i}^{k}(x)$ as $d_{i}^{k}$, we obviously obtain

\begin{equation} \label{haarana} 
 s_{i}^{k-1}  =  \frac{1}{2} s_{2 i}^{k} + \frac{1}{2} s_{2 i+1}^{k} 
\hspace{1cm} ; \hspace{1cm} 
 d_{i}^{k-1}  =  \frac{1}{2} s_{2 i}^{k} - \frac{1}{2} s_{2 i+1}^{k}  
\end{equation}

\begin{figure}[th]      
     \begin{center}
      \setlength{\unitlength}{1cm}
       \begin{picture}( 8.,2.5)           
        \put(-1.,-.5){\includegraphics{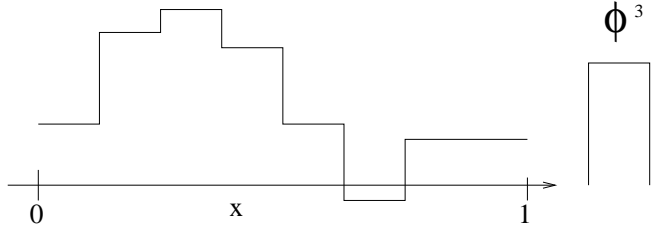}}   
       \end{picture}
       \caption{\label{fhaarm3} The function from Figure~\ref{fhaarm4} at resolution level 3.}
      \end{center}
     \end{figure}

So to calculate the expansion coefficients with respect to the scaling functions 
at the next coarser level, we have to take an average over expansion coefficients 
at the higher resolution level. Because we have to take some weighted sum these 
coefficients are denoted by $s$. To get the expansion coefficients with respect to 
the wavelet, we have to take some weighted difference and the coefficients 
are accordingly denoted by $d$. The wavelet part contains mainly high frequency 
components and by doing this transformation we therefore peel off the highly 
oscillatory parts of the function. The remaining part represented by the 
coefficients $s_{i}^{k-1}$ is therefore smoother. 
For the case of our example 
in Figure~\ref{fhaarm4} the remaining scaling function part after one transformation step is 
shown in Figure~\ref{fhaarm3}.

For any data set whose size is a power of 2, we can now apply this transformation 
repeatedly. In each step the number 
of $s$ coefficients will be cut into half. So we have to stop the procedure as soon  
as there is only one $s$ coefficient left. Such a series of transformation 
steps is called a forward Haar wavelet transform. The wavelet representation of 
the function in Equation~\ref{hsfcrep} is then

\begin{equation} \label{hwvtrep}
f= s_1^0 \phi_1^{0}(x) + d_1^0 \psi_1^{0}(x) + 
      \sum_{i=1}^{2} d_{i}^{1} \psi_{i}^{1}(x) +
      \sum_{i=1}^{4} d_{i}^{2} \psi_{i}^{2}(x) +
      \sum_{i=1}^{8} d_{i}^{3} \psi_{i}^{3}(x)  \: .
\end{equation}

Note that in both cases we need exactly 16 coefficients to represent the function.
Functional representations of this type will be the focus of this article.

By doing a backward wavelet transform, we can go back to the original expansion 
of Equation~\ref{hsfcrep}. Starting at the lowest resolution level, we have to split up 
each scaling function and wavelet on the coarse level into scaling functions 
at the finer level.

\begin{equation} \label{haarsyn} 
 s_{2 i}^{k+1}  =   s_{i}^{k} + d_{i}^{k} 
\hspace{1cm} ; \hspace{1cm} 
 s_{2 i+1}^{k+1}  =  s_{i}^{k} - d_{i}^{k}
\end{equation}

\section{The concept of Multi-Resolution Analysis}
In the previous sections a very intuitive introduction to wavelet theory was given.
The formal theory behind wavelets is called Multi-Resolution Analysis~\cite{daub} (MRA).  
The reader interested in the formal theory can consult Daubechies book. We will list here 
only a few facts which are useful for numerical work.

A bi-orthogonal wavelet family of degree $m$ is characterized by 4 finite filters
denoted by $h_j$, $\tilde{h}_j$, $g_j$, $\tilde{g}_j$. A filter is just a short vector 
which is used in convolutions. Those filters satisfy 
certain orthogonality and symmetry relations.
Scaling functions and wavelets at a coarse level can be written as 
the linear combinations of scaling functions at a higher resolution 
level. These important relations are called refinement relations.
\begin{eqnarray} 
 \phi(x) &=& \sum_{j=-m}^{m} h_j \: \phi(2 x -j) 
                                  \label{refinement1a} \\
 \psi(x) &=& \sum_{j=-m}^{m} g_j \: \phi(2 x -j) 
                                  \label{refinement1b} \\
 \tilde{\phi}(x) &=& 2 \sum_{j=-m}^{m} \tilde{h}_j \: \tilde{\phi}(2 x -j) 
                                  \label{refinement1c} \\
 \tilde{\psi}(x) &=& 2 \sum_{j=-m}^{m} \tilde{g}_j \: \tilde{\phi}(2 x -j) 
                                  \label{refinement1d}  
\end{eqnarray}

The expansion coefficients at different resolution levels are related by the wavelet 
transform equations. The analysis (forward) transform is given by
\begin{equation}
 s_{i}^{k-1} = \sum_{j=-m}^{m} \tilde{h}_{j} s_{j + 2 i}^{k}  
\hspace{1cm} ; \hspace{1cm} 
 d_{i}^{k-1} = \sum_{j=-m}^{m} \tilde{g}_{j} s_{j + 2 i}^{k}  \label{forward} 
\end{equation}

 and a wavelet synthesis (backward) transform is given by
\begin{equation}
 s_{2 i   }^{k+1} =  \sum_{j=-m/2}^{m/2} h_{2 j  } \: s_{i-j}^{k} + g_{2 j  } \: d_{i-j}^{k} 
\hspace{1cm} ; \hspace{1cm} 
 s_{2 i+1 }^{k+1} =  \sum_{j=-m/2}^{m/2} h_{2 j+1} \: s_{i-j}^{k} + g_{2 j+1} \: d_{i-j}^{k} 
 \label{backward}  
\end{equation}
These two equations are generalizations of Equations~(\ref{haarana}) 
and ~(\ref{haarsyn}) which we derived in 
an intuitive way and with a different normalization convention.

The fundamental functions satisfy the following orthogonality relations
\begin{eqnarray}
 \int \tilde{\phi}_{i}^{k}(x) \phi_{j}^{k}(x) dx &=& \delta_{i,j} \label{orthoffa} \\
 \int \tilde{\psi}_{i}^{k}(x) \phi_{j}^{q}(x) dx &=& 0  \: , \: k  \geq q \label{orthoffb} \\
 \int \psi_{i}^{k}(x) \tilde{\phi}_{j}^{q}(x) dx &=& 0  \: , \: k  \geq q \label{orthoffc} \\
 \int \psi_{i}^{k}(x) \tilde{\psi}_{j}^{q}(x) dx &=& \delta_{k,q}  \delta_{i,j} \label{orthoffd}  
\end{eqnarray}
The scaling function is usually normalized to 1
\begin{equation} \label{sfnrm}
 \int \phi(x) dx = 1 
\end{equation}

\section{The fast wavelet transform}

Let us first look at the forward transform given by Equation~\ref{forward} .
The peeling off of the high frequency components in the forward transform 
can be illustrated in the following way:

\bigskip
\bigskip
     
$s^{4}$  $\rightarrow$  $s^{3}$  $\rightarrow$  $s^{2}$  $\rightarrow$  $s^{1}$  $\rightarrow$  $s^{0}$  

\hspace{.4cm}  $\searrow$ \hspace{.35cm}  $\searrow$ \hspace{.35cm} $\searrow$ \hspace{.35cm}   $\searrow$          

  \hspace{0.9cm} $d^{3}$ \hspace{.45cm}  $d^{2}$   \hspace{.45cm}  $d^{1}$   \hspace{.45cm }$d^{ 0}$  

\bigskip
\bigskip

We note that just two arrays of length n (where n is a power of 2) are necessary 
to do the transform as shown below:

\bigskip
\bigskip
original data 

Array 1: $s_0^{4} \: $ $s_1^{4} \: $ $s_2^{4} \: $ $s_3^{4} \: $ $s_4^{4} \: $ $s_5^{4} \: $ $s_6^{4} \: $ $s_7^{4} \: $ $s_8^{4} \: $ $s_9^{4} \: $ $s_{10}^{4}  $ $s_{11}^{4}  $ $s_{12}^{4}  $ $s_{13}^{4}  $ $s_{14}^{4}  $ $s_{15}^{4}  $ 

\bigskip
after first sweep 

Array 2: $s_0^{3} \: $ $s_1^{3} \: $ $s_2^{3} \: $ $s_3^{3} \: $ $s_4^{3} \: $ $s_5^{3} \: $ $s_6^{3} \: $ $s_7^{3} \: $ $d_0^{3} \: $ $d_1^{3} \: $ $d_2^{3} \: $ $d_3^{3} \: $ $d_4^{3} \: $ $d_5^{3} \: $ $d_6^{3} \: $ $d_7^{3} \: $ 

\bigskip
after second sweep 

Array 1: $s_0^{2} \: $ $s_1^{2} \: $ $s_2^{2} \: $ $s_3^{2} \: $ $d_0^{2} \: $ $d_1^{2} \: $ $d_2^{2} \: $ $d_3^{2} \: $ $d_0^{3} \: $ $d_1^{3} \: $ $d_2^{3} \: $ $d_3^{3} \: $ $d_4^{3} \: $ $d_5^{3} \: $ $d_6^{3} \: $ $d_7^{3} \: $ 

\bigskip
after third sweep 

Array 2: $s_0^{1} \: $ $s_1^{1} \: $ $d_0^{1} \: $ $d_1^{1} \: $ $d_0^{2} \: $ $d_1^{2} \: $ $d_2^{2} \: $ $d_3^{2} \: $ $d_0^{3} \: $ $d_1^{3} \: $ $d_2^{3} \: $ $d_3^{3} \: $ $d_4^{3} \: $ $d_5^{3} \: $ $d_6^{3} \: $ $d_7^{3} \: $ 

\bigskip
final data 

Array 1: $s_0^{0} \: $ $d_0^{0} \: $ $d_0^{1} \: $ $d_1^{1} \: $ $d_0^{2} \: $ $d_1^{2} \: $ $d_2^{2} \: $ $d_3^{2} \: $ $d_0^{3} \: $ $d_1^{3} \: $ $d_2^{3} \: $ $d_3^{3} \: $ $d_4^{3} \: $ $d_5^{3} \: $ $d_6^{3} \: $ $d_7^{3} \: $ 

\bigskip
\bigskip

Note that this transformation from the "original data" to the "final data" 
corresponds exactly to the transformation done in an intuitive way to get from 
Equation~\ref{hsfcrep} to Equation~\ref{hwvtrep}. 
Just as in the case of a Fast Fourier transform we have $Log_2(n)$ sweeps to do a full 
transform. However in the case of the wavelet transform the active data set 
(the s coefficients) is cut into half in each sweep. If our filters $h$ and $g$ have 
length $2 m$ the operation count is then given by $2 m (n +n/2 + n/4 + ...)$.
Replacing the finite geometric series by its infinite value, the total operation 
count is thus given by  $4 m n$

The backward transform (Equation~\ref{backward})  can 
pictorially be represented by the following diagram:

\bigskip
\bigskip
     
$s^{4}$  $\leftarrow$  $s^{3}$  $\leftarrow$  $s^{2}$  $\leftarrow$  $s^{1}$  $\leftarrow$  $s^{0}$  

\hspace{.4cm}  $\nwarrow$ \hspace{.35cm}  $\nwarrow$ \hspace{.35cm} $\nwarrow$ \hspace{.35cm}   $\nwarrow$          

\hspace{0.9cm} $d^{3}$  \hspace{.45cm}  $d^{2}$   \hspace{.45cm}  $d^{1}$   \hspace{.45cm }$d^{ 0}$  

\bigskip
\bigskip

As can easily been seen the operation count is again $4 m n$ and again it can be 
done with 2 arrays of length n.
Since each sweep in a wavelet transform is a linear operation it 
can be represented by a matrix.  Denoting the matrix for one sweep in a 
forward transform by $\tilde{F}$ and in a backward transform 
by $B$ we have
\begin{equation}
 F^{T}  =  \tilde{F}^{-1} = B  
\hspace{1cm} ; \hspace{1cm} 
 \tilde{B}^{T}  =  B^{-1} = \tilde{F} \label{transmat} 
\end{equation}
where the tilde on the matrix means that the filter coefficients necessary to fill 
the matrix are replaced by their dual counterparts. Obviously all these matrices are 
sparse and banded.

Backward wavelet transforms can also be used to make plots of scaling functions 
and wavelets.
To generate the scaling function we start with 
a data set where $s_0^{0}=1$ and $d_i^{k}=0$ for all possible $i$'s and $k$'s up 
to a maximum resolution level $k=K$. In the wavelet case the initial data 
set is $s_0^{0}=0$, $d_1^{0}=1$, and $d_i^{k}=0$ for all other values of $i$ and $k$ 
up to the maximal resolution $K$. By doing repeated backward transform sweeps, we 
express these two functions by skinnier and skinnier scaling functions and the 
$s$ coefficients will finally be the functional values within the resolution of the eye.

\section{Interpolating wavelets}
In addition to being advantageous as basis sets, interpolating wavelets 
are also conceptually the simplest wavelets and we will therefore briefly describe 
their construction.
The construction of interpolating wavelets is closely connected to the 
question of how to construct a continuous function $f(x)$ if only its values $f_i$ on 
a finite number of grid points $i$ are known. One way to do this is by recursive 
interpolation. In a first step we interpolate the functional values 
on all the midpoints by using for instance the values of two grid 
points to the right and of two grid points to the left of the midpoint. These four functional 
values actually allow us to construct a third order polynomial and we can then 
evaluate it at the midpoint. In the next step, we take this new data set, 
which is now twice as large as the original one, as the input for a new midpoint 
interpolation procedure. This can be done recursively 
ad infinitum until we have a quasi continuous function.

Let us now show, how this interpolation prescription leads to a set of basis functions.
Denoting by the Kronecker $\delta_{i-j}$ a data set which has a nonzero entry only at the 
j-th position, we can write any initial data set also as a linear combination of 
such Kronecker data sets: $f_i = \sum_j f_j \delta_{i-j}$.
Now the whole interpolation procedure is clearly linear, i.e. the sum of two interpolated 
values of two functions is equal to the interpolated value of the sum of these two functions. 
This means that we can instead 
also take all the Kronecker data sets as the input for separate interpolation procedures, 
to obtain a set of functions $\phi(x-j)$. The final interpolated function is
then identical to 

\begin{equation} \label{polcof}
f(x)= \sum_j f_j \phi(x-j)
\end{equation}

If the initial grid values $f_i$ were 
the functional values of a polynomial of degree less than four, we obviously will have 
exactly reconstructed the original function from its values on the grid points. 
Since any smooth function can locally be well approximated by a polynomial, these 
functions $\phi(x)$ are good basis functions also in the case where $f$ is not 
a polynomial and we will use them as scaling functions to construct a wavelet family. 

The first construction steps of an interpolating scaling function are shown below 
for the case of linear interpolation. The initial Kronecker data set is denoted by 
the big dots. The additional data points obtained after the first interpolation step 
are denoted by medium size dots and the additional data points obtained after the
second step by small dots.

   \begin{figure}[h]    
     \begin{center}
      \setlength{\unitlength}{1cm}
       \begin{picture}( 8.,2.)           
        \put(-1.,0.){\includegraphics{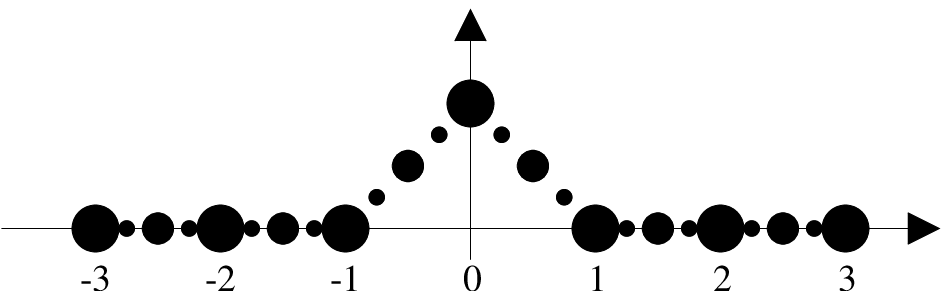}}   
       \end{picture}
      \end{center}
     \end{figure}

\begin{figure}[ht]    
     \begin{center}
      \setlength{\unitlength}{1cm}
       \begin{picture}( 8.,3.5)           
        \put(-4.,-1.5){\includegraphics{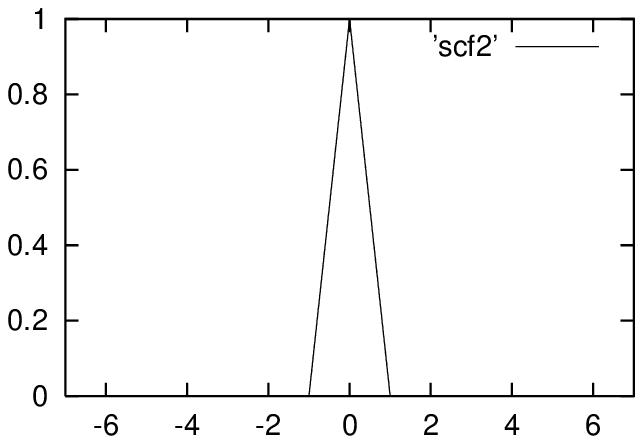}}   
        \put(3.,-1.5){\includegraphics{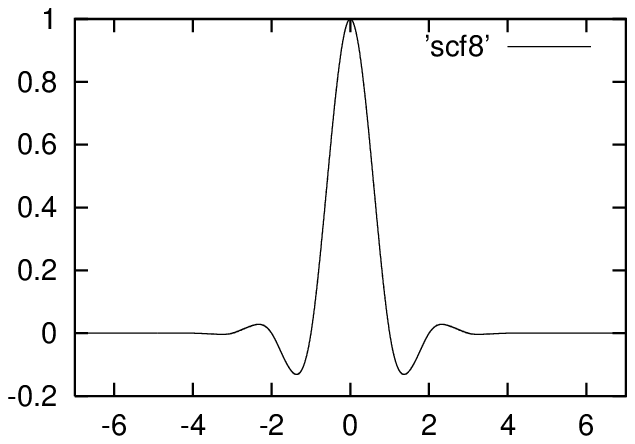}}   
       \end{picture}
       \caption{  \label{scf28} A Kronecker delta interpolated ad infinitum with 
                linear interpolation (left panel) an 7-th order 
                interpolation (right panel) .}
      \end{center}
     \end{figure}

Continuing this process ad infinitum will then result 
in the function shown in the left panel of Figure~\ref{scf28}. If an higher order 
interpolation scheme is used the function shown in the right panel of Figure~\ref{scf28} 
is obtained.

By construction it is clear, that $\phi(x)$ has compact support. If an $(m-1)$-th order 
interpolation scheme is used, the filter length is $(m-1)$ and 
the support interval of the scaling function is $[-(m-1);(m-1)]$.

It is also not difficult to see that the functions $\phi(x)$ satisfy the refinement 
relation. Let us again consider the interpolation ad infinitum of a Kronecker data 
set which has everywhere zero entries except at the origin. We can now split up 
this process into the first step, where we calculate the half-integer 
grid point values, and a remaining series of separate ad infinitum interpolations for 
all half-integer Kronecker data sets, which are necessary to represent the 
data set obtained by the first step. Doing the ad-infinitum interpolation 
for a half integer Kronecker data set with a unit entry at position j, 
we obviously obtain the same scaling function, just 
compressed by a factor of 2, $\phi(2 x - j)$.  
If we are using a $(m-1)$-th order interpolation scheme (i.e. $m$ input data for the 
interpolation process) we thus get the relation

\begin{equation}
 \phi(x) = \sum_{j=-m+1}^{m-1}  \phi(j/2) \; \phi(2 x - j) \label{selfinter} 
\end{equation}

Comparing this equation with the refinement relation Equation~\ref{refinement1a}  
we can identify the first filter $h$ as

$$ h_j = \phi(j/2) \: \: , \: \: j=-m+1,m-1 $$

For the case of third order interpolation the numerical values of $h$  follow 
from the standard interpolation formula and are given by
        $\{$ -1/16 , 0 , 9/16 , 1 , 9/16 , 0 , -1/16 $\}$.

Let us next determine the filter $\tilde{h}$. Let us consider a function $f(x)$ which is 
band-limited in the wavelet sense, i.e which can exactly be represented by a 
superposition of scaling functions at a certain resolution level $K$.

$$ f(x) =  \sum_j s^K_j \phi_j^K(x) $$

It then follows from the orthogonality relation Equation~\ref{orthoffa} that 

\begin{equation}
 s_j^{K} = \int \tilde{\phi}_j^{K}(x) \: f(x)  dx \label{sinteg}
\end{equation}

Now we have seen above that with respect to interpolating scaling functions, 
a band-limited function is just any polynomial of degree less than or equal to $m-1$,
and that in this case the expansion coefficients $s^K_j$ are just the functional values 
at the grid points (Equation~\ref{polcof}). We therefore have 

\begin{equation}
 s_j^{K} = \int \tilde{\phi}_j^K(x) f(x) dx = f_j  \label{polycof}
\end{equation}

which shows that the dual scaling function $\tilde{\phi}$ is the delta function.
\begin{equation} \label{delta1}
\tilde{\phi}(x) = \delta(x)
\end{equation}
Obviously the delta function satisfies a trivial refinement relation 
$\delta(x)= 2 \delta(2x)$ and 
from Equation~\ref{refinement1c} we conclude that  $\tilde{h}_j = \delta_j$
From the symmetry relations for the filters the two remaining filters   $\tilde{g}(i)$ and $g(i)$
can be determined and we have thus completely specified our wavelet family.

Using these filters we can then determine the wavelet $\psi$ and its dual 
counterpart $\tilde{\psi}$ which 
turn out to be
\begin{equation} \label{ipowvlt}
\psi(x)  =  \phi(2 x -1) 
\end{equation}
\begin{equation} \label{delta3}
\tilde{\psi}(x)  = 
 \frac{-1}{16} \delta((x-\frac{1}{2})-3) +
 \frac{ 9}{16} \delta((x-\frac{1}{2})-1) -
 \delta((x-\frac{1}{2})) +
 \frac{ 9}{16} \delta((x-\frac{1}{2})+1) +
 \frac{-1}{16} \delta((x-\frac{1}{2})+3) 
\end{equation}

We see that the interpolating wavelet is a very special case in that its 
scaling function and wavelet have the same functional form and that the dual 
functions are related to the delta function. The non-dual functions are 
shown in Figure~\ref{lazy4}. 

Lifting~\cite{sgwvlt} is a very useful technique to modify an existing family of wavelets to meet 
specific needs. We can for instance lift the interpolating wavelets to obtain a new 
family whose wavelet has more vanishing moments $M_l$.

$$ M_l = \int \psi_j^{K}(x) \: x^l dx  $$

which will for instance improve the frequency properties of the wavelet.

\section{Expanding functions in a wavelet basis}
As was demonstrated in the case of the Haar wavelet, 
there are two possible representations of a function within the framework of 
wavelet theory. The first one is called scaling function representation and 
involves only scaling functions. The second is called wavelet representation
and involves wavelets as well as scaling functions.
Both representations are completely equivalent and exactly the same number of 
coefficients are needed in the case where one has uniform resolution.

The scaling function representation is given by

\begin{equation}
 f(x) = \sum_j s_j^{Kmax} \phi_j^{Kmax}(x) \label{scfrep}
\end{equation}

The coefficients $s_j^{Kmax}$ can be calculated 
by integration through Equation~\ref{polycof}. 
Once we have a set of coefficients $ s_j^{Kmax} $ we can use a full forward wavelet transform 
to obtain the coefficients of the wavelet representation

\begin{equation}
 f(x) = 
        \sum_j s_j^{Kmin} \: \phi_j^{Kmin}(x)  
       + \sum_{K=Kmin}^{Kmax} \sum_j d_j^{K} \; \psi_j^{K}(x)  \label{wvltrep}
\end{equation}

Alternatively, one could also directly calculate the $d$ coefficients by integration

\begin{equation}
 d_j^{K} = \int \tilde{\psi}_j^{K}(x) \: f(x) dx  \label{dinteg}
\end{equation}

Equation \ref{dinteg} follows from the orthogonality relations 
\ref{orthoffb} to \ref{orthoffd}.

So we see that if we want to expand a function either in scaling functions or wavelets, 
we have to perform integrations at some point to calculate the coefficients.
For general wavelet families this integration can be fairly cumbersome~\cite{map} and 
require especially in 2 and 3 dimensions a substantial number of integration 
points. Furthermore it is not obvious how to do the integration if the 
function is only given in tabulated form. 
The interpolating wavelets discussed above are the glorious exception.
Since the dual scaling function is a delta function~(\ref{delta1} )
and since the dual wavelet is a sum of delta functions~(\ref{delta3} ), 
one or a few data points are sufficient to do the integration exactly. 
One will therefore 
get exactly the same number of coefficients as one has data points  
and one has an invertible 
one-to-one mapping between the functional values on the grid and the expansion 
coefficients.
This is even true in the case of nonuniform data sets, where we  necessarily 
have to calculate the $s$ and $d$ coefficients directly by integration using~\ref{dinteg}.
As follows 
from Equation~\ref{delta1} and~\ref{delta3}, one just needs 
the functional values at the data point at which the wavelet will be centered and 
a few data points at one lower resolution level around this center. If one 
wants to calculate the interpolating wavelet center at the high resolution grid point 
indicated by the fat arrow in the figure below, one needs in the 
case of the 4-th order interpolating wavelets the 4 additional points indicated by thin arrows 
which belong to a more coarse grid and are therefore always available even if the fine 
grid does not extend into this region.

   \begin{figure}[h]  
     \begin{center}
      \setlength{\unitlength}{1cm}
       \begin{picture}( 8.,1.5)           
        \put(-3.,-0.){\includegraphics{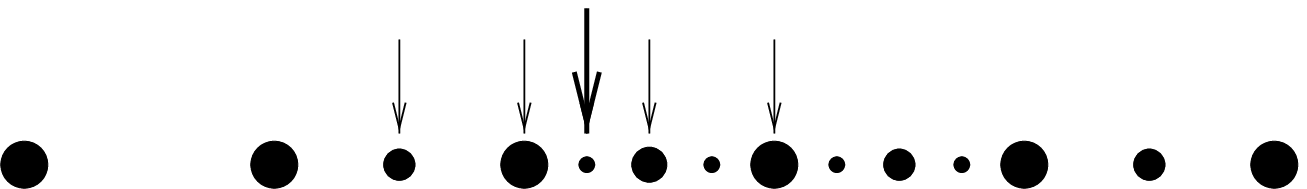}}   
       \end{picture}
      \end{center}
     \end{figure}

In the case where one wants to represent functions with several length scales which need 
inhomogeneous real space grid structure the wavelet representation allows 
a much more compact representation than the scaling function representation, since on 
can neglect all the tiny $d$ coefficients in the regions where one has little variation.
To illustrate this let us look at the function $f$
$$ f(x) = \sum_{l=1}^{8} exp(-(x l)^2) $$
Evidently this function exhibits 8 different length scales.
If one expands one simple Gaussian $exp(-x^2)$ with respect to 4-th order 
interpolating scaling functions
with a resolution of 1/16, one gets a reasonably small error of $10^{-6}$. 
For the multi-scale function $f$, this error increases 
to more than $10^{-2}$ with the same resolution. If one however uses 
a scheme where one uses 32 wavelets on additional 5 resolution levels to improve 
the resolution as one approaches the origin 
one can again represent the function with an error of roughly 
$10^{-6}$ ( it turns out that the expected 8 additional levels are not all needed).
The total number of coefficients needed to represent the function 
in the interval $[-2;2]$ is then 4 $\times$ 16 coefficients 
for the equal resolution (1/16) 
scaling function part plus 5 $\times$ 32 coefficients for the resolution enhancement 
with the wavelets, which makes all together 224 coefficients. This has to be compared 
with the 1024 scaling function coefficients which would be needed to represent 
the function over the whole interval with the maximum resolution of (1/256), which 
we have obtained around the origin with this data compression scheme.

\section{Wavelets in 2 and 3 dimensions}
The easiest way to construct a wavelet basis in higher dimensional spaces is 
by forming product functions~\cite{daub}. For simplicity of notation we will only 
consider here the 2-dimensional case, the generalization to 
higher dimensional spaces being obvious.

The space of all scaling functions of resolution level $k$ is given by

\begin{equation} \label{ss}
\phi_{i1,i2}^{k}(x,y) = \phi_{i1}^{k}(x) \phi_{i2}^{k}(y)
\end{equation}
The wavelets consist of three types of products
\begin{eqnarray}
 \psi[sd]_{i1,i2}^{k}(x,y) =  \phi_{i1}^{k}(x) \psi_{i2}^{k}(y) \label{sd}  \\
 \psi[ds]_{i1,i2}^{k}(x,y) =  \psi_{i1}^{k}(x) \phi_{i2}^{k}(y) \label{ds}  \\
 \psi[dd]_{i1,i2}^{k}(x,y) =  \psi_{i1}^{k}(x) \psi_{i2}^{k}(y) \label{dd}
\end{eqnarray}

A wavelet transform step in the 2-dimensional 
setting is done by first transforming along the x and then along the y 
direction (or vice versa).

\section{The standard operator form}
In a bi-orthogonal wavelet basis it is natural to solve a differential equation 
in the collocation sense. Let us recall that in the collocation method one has 
two functional spaces, the space of the basis function which are used 
to represent the solution and the space of the test functions which are used 
to multiply the differential equation from the left to obtain a linear system 
of equations.  In our case the expansion set are the scaling functions and 
wavelets while the test set are their dual counterparts. 
Lets consider the case of Poisson's equation

\begin{equation}
 \nabla^2 V = -4 \pi \rho . \label{poisson}
\end{equation}

\begin{figure}[ht]        
     \begin{center}
      \setlength{\unitlength}{1cm}
       \begin{picture}( 6.,6.5)           
        \put(-1.,-.2){\includegraphics{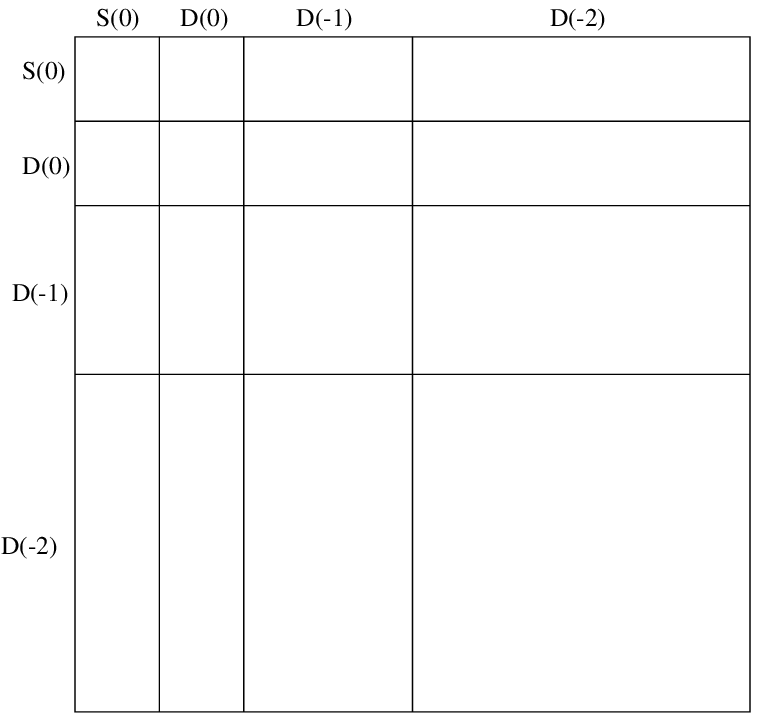}}   
       \end{picture}
       \caption{\label{standard} The structure of a matrix in the standard form.}
      \end{center}
     \end{figure}

Given the expansion of the charge density $\rho$ in a wavelet basis

\begin{equation}
 \rho(x) = 
           \sum_j s_j^{Kmin} \: \phi_j^{Kmin}(x)  \label{rhoex}
        +  \sum_{K=Kmin}^{Kmax} \sum_j d_j^{K} \; \psi_j^{K}(x)  
\end{equation}

we are looking for the wavelet expansion coefficients of the potential $V$.
\begin{equation}
 V(x) = 
          \sum_j S_j^{Kmin} \: \phi_j^{Kmin}(x)  \label{vex}
       +  \sum_{K=Kmin}^{Kmax} \sum_j D_j^{K} \; \psi_j^{K}(x)  
\end{equation}
Plugging in the expansion for $\rho$ and $V$ (~\ref{rhoex}) and (~\ref{vex}) in Poissons
equation ~\ref{poisson} and
multiplying from the left with the dual wavelet collocation test space we obtain a system of 
equations

\begin{equation}
 A_s \vec{v} = \vec{\rho} 
\end{equation}

where $\vec{v}$ is the vector containing both the $s$ and $d$ coefficients of the 
potential and $\vec{\rho}$ is the same vector for the charge density $\rho$.
The matrix $A_s$ represents the Laplacian in this wavelet basis and one says
that it has standard form. This standard form is graphically shown in Figure~\ref{standard} .

The problem with the standard form is that it is first of all rather complicated.
There is coupling between all resolution levels and one has to calculate 
many different types of matrix elements corresponding to all possible 
products of wavelets and scaling functions at different resolution levels and 
positions. The second point is that there are many blocks in that matrix which 
have no or only few zeroes. Let us look at the blocks representing the 
coupling between the scaling functions at the highest resolution level 
and the wavelets at the different resolution. In general each scaling function will 
extend over the whole computational volume and will therefore overlap 
with all the wavelets at any position. All these blocks will consequently have 
nonzero entries only. So this standard matrix form has more nonzero 
entries than we would like to have for optimal efficiency in the matrix vector 
multiplications which are required for all iterative linear equation solvers.

\section{The non-standard operator form}
The so-called nonstandard~\cite{nonstand} form gives a much easier and efficient representation 
of our matrix. To derive it let us first assume, that our potential $V$ and 
charge $\rho$ are given in a scaling function basis.
The Laplacian is then represented by a matrix $A$ whose elements $A_{i,j}$ are given by 
$\int \tilde{\phi}^k_i(x) \nabla^2 \phi^k_j(x) dx$

The matrix equation 
$$ A \vec{v} = \vec{\rho} $$ 
can graphically be represented in the following way:

   \begin{figure}[ht]
     \begin{center}
      \setlength{\unitlength}{1cm}
       \begin{picture}( 6.,1.)           
        \put(-0.,-0.5){\includegraphics{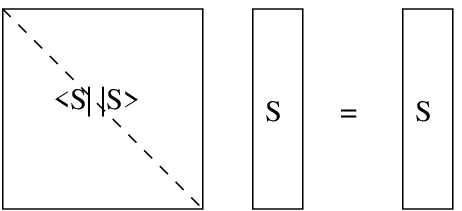}}   
       \end{picture}
      \end{center}
     \end{figure}

Now we can of course perform one 
step of a forward wavelet transform  on all our data, i.e. both on the vector 
to be multiplied with the matrix and on the vector which is the result of this 
matrix times vector multiplication. Correspondingly we have then to transform 
the matrix $A$ using the matrices whose properties are given in Equation~\ref{transmat}.

$$ \tilde{F} \vec{\rho} = (\tilde{F} A F^T) (\tilde{F} \vec{v}) $$
Graphically this gives:

   \begin{figure}[ht]
     \begin{center}
      \setlength{\unitlength}{1cm}
       \begin{picture}( 6.,1.0)           
        \put(-0.,-0.5){\includegraphics{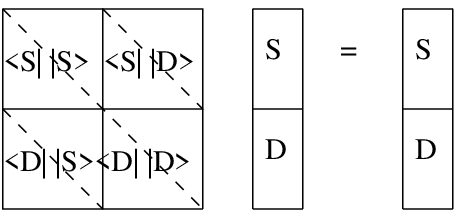}}   
       \end{picture}
      \end{center}
     \end{figure}

If we recursively applied wavelet transform to the upper $S$ part we would obviously 
obtain the standard operator form. To get the nonstandard form, we have to add another 
step where we artificially enlarge our matrix $A$ by putting in 5 blocks of zeroes
as shown below:

   \begin{figure}[ht]
     \begin{center}
      \setlength{\unitlength}{1cm}
       \begin{picture}( 6.,2.3)           
        \put(-0.3,-0.7){\includegraphics{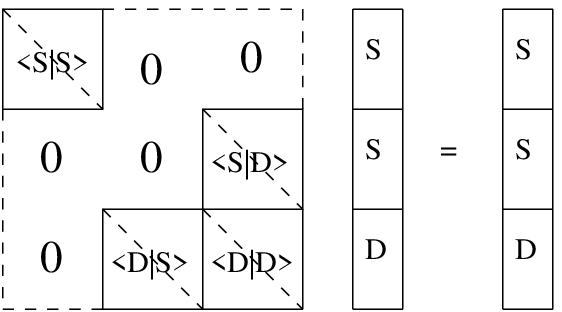}}   
       \end{picture}
      \end{center}
     \end{figure}

We see that our input and output vectors $\vec{v}$ and $\vec{\rho}$ also have to be adapted to this 
matrix structure leading to a redundant copy of the $S$ data set.

\begin{figure}[ht]        
     \begin{center}
      \setlength{\unitlength}{1cm}
       \begin{picture}( 10.0,9.)           
        \put(-1.,-0.0){\includegraphics{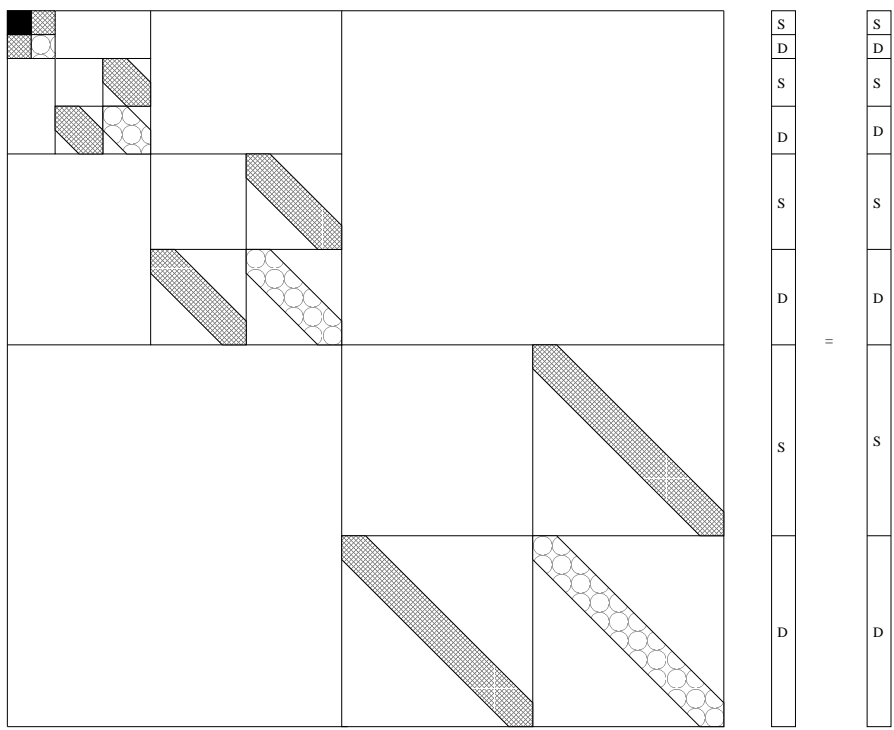}}   
       \end{picture}
       \caption{\label{nonstf} The structure of a matrix in the nonstandard form.}
      \end{center}
     \end{figure}

We can now recursively apply this 2-step procedure on the $<S|S>$ block 
of the resulting matrices. Doing this we obtain the so called non-standard form, 
which is graphically visualized in Figure~\ref{nonstf} 

As we see, we have now completely decoupled different resolution levels, since 
there are no blocks in this matrix between different levels. The coupling between 
different levels just enters trough the wavelet transforms which have to be interleaved 
with the application of this nonstandard operator form. We also see that all the 
nonzero blocks of this nonstandard matrix representation are strictly banded and 
the application of this matrix to a vector scales therefore linearly.

The structure of the matrix in Figure~\ref{nonstf} is primarily valid for the case 
of uniform resolution where all the possible d coefficients at the 
highest resolution level are nonzero. It can however easily be seen that 
this nonstandard form retains its advantage in a case of varying resolution 
where only some of the d coefficients are nonzero. 
If the nonredundant input data set is sparse, the redundant input data set 
will be sparse as well. Since all the blocks are banded, the redundant output 
set will be sparse as well. Finally the nonredundant output set will 
then be sparse as well.

\section{Calculation of differential operators in a wavelet basis}
\label{caldifop}
As we have seen in the preceeding chapter we need the matrix elements 

\begin{eqnarray}
 \int  \tilde{\phi}_{i}^{k}(x) \frac{\partial^l}{\partial x^l} \phi_{j}^{k}(x) & dx & \\
 \int  \tilde{\psi}_{i}^{k}(x) \frac{\partial^l}{\partial x^l} \phi_{j}^{k}(x) & dx & \\
 \int  \tilde{\phi}_{i}^{k}(x) \frac{\partial^l}{\partial x^l} \psi_{j}^{k}(x) & dx & \\
 \int  \tilde{\psi}_{i}^{k}(x) \frac{\partial^l}{\partial x^l} \psi_{j}^{k}(x) & dx &  
\end{eqnarray}

for the application of an operator in the nonstandard form. 
Matrix elements on different resolution levels are related by simple 
scaling relations. 
So we just have to calculate these 4 matrix elements for one resolution level.
On a certain resolution level, we can use the  refinement relations to express 
the  matrix elements involving wavelets in terms of matrix elements involving 
scaling functions (at a better resolution level) only.
So we just have to calculate the basic integral $a_i$

\begin{equation}
a_i =  \int  \tilde{\phi}(x) \frac{\partial^l}{\partial x^l} \phi(x-i) dx  
\end{equation}

Using the refinement relations Equations~\ref{refinement1a} and ~\ref{refinement1c} for 
$\phi$ and $\tilde{\phi}$ we obtain

\begin{eqnarray}
a_i & = &  \int  \tilde{\phi}(x) \frac{\partial^l}{\partial x^l} \phi(x-i) dx   \\
& = & \sum_{\nu,\mu} 2 \tilde{h}_{\nu} h_{\mu} \int 
      \tilde{\phi}(2x-\nu) \frac{\partial^l}{\partial x^l} \phi(2x-2i-\mu) dx \nonumber \\
& = & \sum_{\nu,\mu} 2 \tilde{h}_{\nu} h_{\mu} 2^{l-1} \int 
      \tilde{\phi}(y-\nu) \frac{\partial^l}{\partial y^l} \phi(y-2i-\mu) dy \nonumber \\
& = & \sum_{\nu,\mu} \tilde{h}_{\nu} h_{\mu} 2^l \int 
      \tilde{\phi}(y) \frac{\partial^l}{\partial y^l} \phi(y-2i-\mu+\nu) dy \nonumber \\
& = & \sum_{\nu,\mu} \tilde{h}_{\nu} h_{\mu} \: 2^l \: a_{2i-\nu+\mu}
\end{eqnarray}

We thus have to find the eigenvector $\vec{a}$ associated with the eigenvalue of 
$2^{-l}$.

\begin{equation} \label{eigsys}
\sum_j A_{i,j} \: a_j = \left( \frac{1}{2} \right) ^l a_i
\end{equation}

where the matrix $A_{i,j}$ is given by

\begin{equation} 
A_{i,j} = \sum_{\nu,\mu} \tilde{h}_{\nu} h_{\mu} \: \delta_{j,2i-\nu+\mu}
\end{equation}

As it stands this eigensystem has a solution only if the 
rang of the matrix $A-2^{-l}I$ is less than its dimension. For a well defined 
differential operator, i.e if $l$ is less than the degree of smoothness of the 
scaling function this will be the case. 

The system of equations~\ref{eigsys}  determines the $a_j$'s only 
up to a normalization factor. For the case of interpolating wavelets the 
normalization condition is easily found from the requirement that one obtains the 
correct result for the function $x^l$.
From the normalization of the scaling function~(\ref{sfnrm}) and from elementary calculus, it 
follows that 

\begin{equation} \label{norm1} 
\int \phi(x) \frac{\partial^l}{\partial x^l} x^l dx = \int \phi(x) l! dx = l!
\end{equation}

On the other hand we know, that we can expand any polynomial of low enough degree 
exactly with the interpolating polynomials. The expansion coefficients are 
just $i^l$ by Equation~\ref{polycof}. So we obtain

\begin{equation} \label{norm2}
\int \phi(x) \frac{\partial^l}{\partial x^l} \sum_i i^l \phi(x-i) = \sum_i i^l a_i
\end{equation}

By comparing Equation~\ref{norm1}  and ~\ref{norm2} we thus obtain 
the normalization condition

\begin{equation} \label{normd}
\sum_i i^l a_i = l!
\end{equation}

The interpolating wavelet family offers also an important advantage for the calculation of 
differential operators. Whereas in general derivative filters extend 
over the interval $[-2m ; 2m]$ their 
effective filter length is only $[-m+2 ; m-2]$. Since higher-dimensional wavelets are 
products of one-dimensional ones differential operators in the higher-dimensional case 
can easily be derived from the one-dimensional results.

The standard operator form can not only be used for the application of differential 
operators, but also for other operations. If one want to transform for instance from 
one wavelet family $\phi$ to another wavelet family $\Phi$ the basic integral becomes

\begin{equation}
a_i  =   \int  \tilde{\Phi}(x)  \phi(x-i) dx  =  
   \sum_{\nu,\mu} \tilde{H}_{\nu} h_{\mu} a_{2i-\nu+\mu}
\end{equation}

Another use is for scalar products where the fundamental integral is 

\begin{equation}
a_i  =   \int  \phi(x)  \phi(x-i) dx  =  \frac{1}{2} 
  \sum_{\nu,\mu} h_{\nu} h_{\mu} a_{2i-\nu+\mu}
\end{equation}

\section{Solving Poisson's equation for the $U_2$ dimer}
Poisson's equation is a prototype differential equation and we want to solve it 
therefore as an illustration of wavelet theory.
To demonstrate the power of the wavelet method we applied it to the 
most difficult system we could think of in the area of electronic structure 
calculations, namely the calculation of the electrostatic potential of 
a three dimensional $U_2$ dimer~\cite{sgoi}.
In this example, we clearly find widely varying length scales. The valence 
electrons have an extension of 5 atomic units, the 1s core electrons of 
2/100 atomic units and the nucleus itself was represented by a charge distribution 
with an extension of 1/2000 atomic units. So all together the length scales 
varied by 4 orders of magnitude and two regions of increasing resolution 
(around each nucleus) were needed. In order to have quasi perfect natural boundary 
conditions we embedded the molecule in a computational volume of side length 
$10^4$ atomic units. All together this necessitated 22 levels of resolution.
Even though the potential itself varies by many orders of magnitude, we were 
able to calculate the solution with typically 7 digits of accuracy.
We believe that it would not be possible with any other method to 
solve this kind of benchmark problem.

\begin{figure}[ht]        
     \begin{center}
      \setlength{\unitlength}{1cm}
       \begin{picture}( 8.,3.0)           
        \put(-0.,.0){\includegraphics{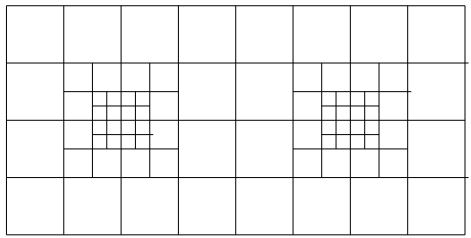}}   
       \end{picture}
       \caption{\label{grid2} A grid with two centers of increased resolution around 
                the two nuclei. Only 3 of the 22 levels used in the calculation are 
                shown in this projection on a plane.}
      \end{center}
     \end{figure}

The solution of Poisson's equation consists of several steps.
Initially we have to find the wavelet expansion for a data set on a nonuniform real 
space grid structure shown in Figure~\ref{grid2} which represents the charge density.
The resolution needed can in this example be estimated from the known extension 
and variation of the different atomic shells.
Analogously to the one-dimensional case, this expansion can also easily be obtained for 
higher dimensional interpolating wavelets since all the dual function are related to 
delta functions. Let us point out, that also in this case the 
mapping from real space representation to the wavelet representation is invertible, and 
we could thus get back exactly the same real space values if we evaluate the 
wavelet expansion on the grid points.

\begin{figure}[ht]    
     \begin{center}
      \setlength{\unitlength}{1cm}
       \begin{picture}( 6.,4.5)           
        \put(-5.7,-1.5){\includegraphics{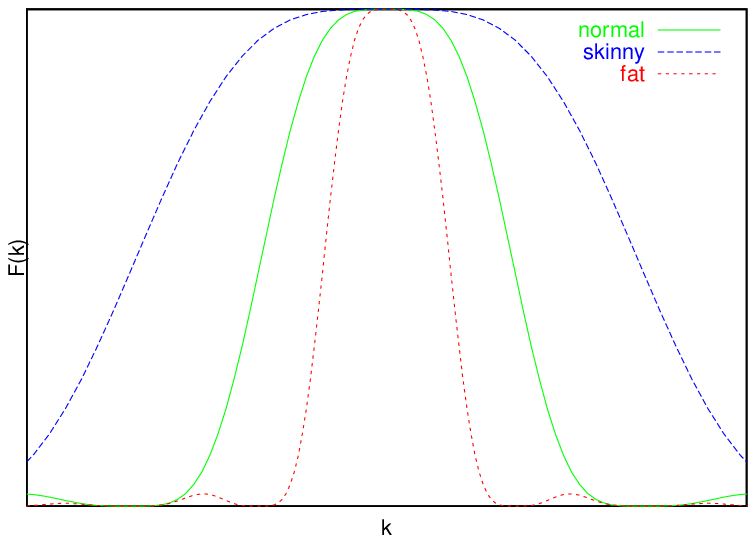}}   
        \put( 2.3,-1.5){\includegraphics{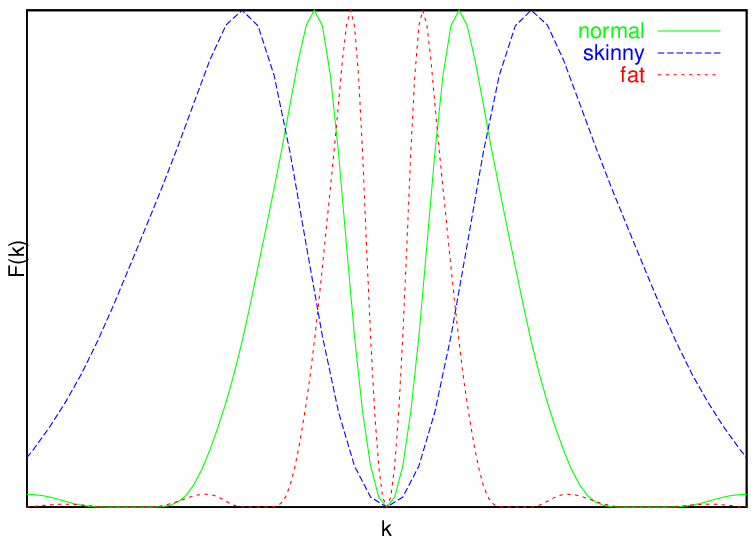}}   
       \end{picture}
       \caption{\label{freqloc} The Fourier spectrum of a 4-th order 
                unlifted Lazy wavelet (left panel) and lifted wavelet (right panel).
                The spectrum is shown for 3 wavelets on neighboring 
                resolution levels. One has reasonable 
                frequency separation in the lifted but not in the unlifted case. }
      \end{center}
     \end{figure}

Next we start a iteration loop for the potential.
First we have to apply the Laplace operator to an approximate 
potential using the non-standard operator form. Subtracting from this result the 
charge density gives the residue vector which is the basis for all iterative methods~\cite{recipes}, 
such as steepest descent and conjugate gradient methods. Unfortunately the condition number 
of the Laplace matrix worsens when more high resolution levels are added and the number of 
iterations needed to obtain convergence would dramatically increase if we used straightforward 
iterative methods. It this therefore absolutely necessary to use a preconditioned iterative 
method which will give a condition number which is independent of the maximal resolution.
In a preconditioning scheme one has to find an approximate inverse matrix of the Laplace matrix.
If the Laplace matrix is strongly diagonally dominant, then just the inverse of the diagonal 
part (which is again diagonal) will be a good approximate inverse. Whether the Laplace matrix 
is strongly diagonally dominant depends on the kind of wavelet family which is used. 
In a plane wave representation the Laplace matrix is strictly diagonal. If therefore our 
wavelet family has good frequency localization properties the resulting matrix 
will be strongly diagonally dominant. Unfortunately our favorite interpolating wavelets
have a very poor frequency localization making an iterative solution 
practically impossible. It is therefore necessary to do the preconditioning 
step within another family such as the lifted interpolating wavelets which have much better 
frequency localization properties as shown in Figure~\ref{freqloc}. 
Their improved frequence localization properties is related to the fact that 
several moments of the wavelet vanish.

As discussed above the transformation into 
another wavelet family can also be done with the help of the non-standard operator 
form.  The preconditioned residue vector is then used to update the potential and we 
go back to the beginning of the iteration.
Using lifted interpolating wavelets with 2 vanishing moments we were able to reduce 
the norm of the 
residue vector by one order of magnitude with 3 iterations independent of the 
maximal resolution. Despite their poor frequency localization properties,  
unlifted interpolating wavelets have recently also been proposed for the solution of 
Poisson's equation~\cite{lippert}.

\section{Outlook and conclusions}
Since we used mainly interpolating wavelets, all we did was essentially interpolating, 
which is one of the oldest technique in numerical analysis. However the 
framework provided by wavelet theory puts this whole interpolation procedure 
on the new and powerful basis of multi-resolution analysis, expanding thus 
considerably the scope of interpolation based techniques. In particular 
it assigns basis functions to certain interpolation schemes. Wavelet based 
techniques allow us thus to solve differential equations 
which have several length scales and to do this with linear scaling. 
It is thus to be expected, that 
wavelet based techniques will catalyze progress in many fields of science
and engineering, where such problems exist.
An detailed tutorial style book describing how to use wavelets for the solution of 
partial differential equations will soon be published by the authors.

\end{document}